\documentclass[11pt]{article}%
\usepackage{graphicx}
\usepackage{amsmath}
\usepackage{amsfonts}
\usepackage{amssymb}%
\providecommand{\U}[1]{\protect\rule{.1in}{.1in}}
\newtheorem{theorem}{Theorem}[section]

\newtheorem{corollary}[theorem]{Corollary}

\newtheorem{lemma}[theorem]{Lemma}

\newenvironment{proof}[1][Proof]{\textbf{#1.} }{\ \rule{0.5em}{0.5em}}
\large\normalsize
\begin{document}

\title{Finding a Shortest $M$-link Path in a Monge Directed Acyclic Graph}
\author{Joy Z. Wan\\Siebel School of Computing and Data Science\\University of Illinois Urbana-Champaign }
\date{}
\maketitle

\begin{abstract}
A Monge directed acyclic graph (DAG) $G$ on the nodes $1,2,\cdots,N$ has edges
$\left(  i,j\right)  $ for $1\leq i<j\leq N$ carrying submodular edge-lengths.
Finding a shortest $M$-link path from $1$ to $N$ in $G$ for any given
$1<M<N-1$ has many applications. In this paper, we give a contract-and-conquer
algorithm for this problem which runs in $O\left(  \sqrt{NM\left(  N-M\right)
\log\left(  N-M\right)  }\right)  $ time and $O\left(  N\right)  $ space. It
is the first $o\left(  NM\right)  $-time algorithm with linear space
complexity, and its time complexity decreases with $M$ when $M\geq N/2$. In
contrast, all previous strongly polynomial algorithms have running time
growing with $M$. For both $O\left(  poly\left(  \log N\right)  \right)  $ and
$N-O\left(  poly\left(  \log N\right)  \right)  $ regimes of $M$, our
algorithm has running time $O\left(  N\cdot poly\left(  \log N\right)
\right)  $, which partially answers an open question rased in \cite{AST94}
affirmatively. 

\end{abstract}

\bigskip

\section{Introduction}

\bigskip

Many algorithmic problems \cite{AST94,BLP92} can be reduced to seeking a
shortest $M$-link path in a Monge directed acyclic graph (DAG) $G$ on nodes
$1,2,\cdots,N$ with $1\leq M\leq N-1$. The edge set of $G$ is $\left\{
\left(  i,j\right)  :1\leq i<j\leq N\right\}  $; and the edge-length (or
edge-weight) function $c$ of $G$ satisfies that for any $1\leq i<j<k<l\leq
N$,
\[
c\left(  i,l\right)  +c\left(  j,k\right)  \geq c\left(  i,k\right)  +c\left(
j,l\right)  .
\]
In addition, the length $c\left(  i,j\right)  $ of each edge $\left(
i,j\right)  $ can be evaluated in \emph{constant} time. A\ path $P$ in $G$
specified by the sequence
\[
s=v_{0},v_{1},\cdots,v_{m}=t
\]
is said to be an $m$-link $s$-$t$ path, and its length is the total length of
edges in $P$. The path $P$ is called a \emph{shortest} $m$-link $s$-$t$ path
if $P$ has the minimum length among all $m$-link $s$-$t$ paths.

\bigskip

As observed by \cite{AST94,BLP92}, a shortest $M$-link $1$-$N$ path in $G$ can
be computed in $O\left(  NM\right)  $ time and space by the standard dynamic
programming accelerated by the celebrated SMAWK algorithm \cite{AKM+87}. A
parametric search scheme was also introduced in \cite{AST94,BLP92} for
computing a shortest $M$-link $1$-$N$ path in $G$. When the edge lengths are
restricted to integers, the binary search method yields a \emph{weakly
polynomial} algorithm \cite{AST94,BLP92}. A \emph{strongly polynomial}
algorithm with sub-quadratic complexity was developed in \cite{AST94}, which
is referred to the algorithm AST. The algorithm AST follows Megiddo's
parametric search paradigm \cite{Meg83}, and runs in $O\left(  N\sqrt{M\log
N}+N\log N\right)  $ time and $O\left(  N\sqrt{M\log N}\right)  $ space. For
the $\Omega\left(  \log N\right)  $ regime of $M$, the algorithm AST
\cite{AST94} is superior to the $O\left(  NM\right)  $-time accelerated
dynamic programming; and the existence of an $O\left(  N\cdot poly\left(  \log
n\right)  \right)  $-time algorithm was raised as an open question in
\cite{AST94}. 

\bigskip

For the same $\Omega\left(  \log N\right)  $ regime of $M$, Schieber
\cite{Sch98} gave a \emph{recursive} parametric search algorithm with
$O\left(  N2^{4\sqrt{\left(  \log M\right)  \left(  \log\log N\right)  }%
}\right)  $ running time. Schieber's algorithm is superior to the algorithm
AST \cite{AST94} when $M=\Theta\left(  N\right)  $, but may be inferior when
$M=\Theta\left(  poly\left(  \log N\right)  \right)  $. Indeed when
$M=\log^{16}N$, Schieber's algorithm has the same asymptotic time complexity
$O\left(  N\log^{16}N\right)  =O\left(  NM\right)  $ as the accelerated
dynamic programming, while the algorithm AST \cite{AST94} has time complexity
$O\left(  N\log^{8.5}N\right)  $ and is asymptotically faster by a factor of
$\Theta\left(  \log^{7.5}N\right)  $. Similarly, when $M=\log^{4}N$,
Schieber's algorithm has time complexity $O\left(  N\log^{8}N\right)
=O\left(  NM^{2}\right)  $, which is even asymptotically slower by a factor of
$\Theta\left(  M\right)  $ than the accelerated dynamic programming. 

\bigskip

Schieber \cite{Sch98} also noted, with no analysis, the linear space
complexity of his recursive algorithm. A careful look into the recursion
reveals that the space complexity hidden within the recursion is more than
linear order. The recursion runs on an auxiliary DAG with modified lengths of
\emph{all} edges from the node $1$. These modified lengths have to be
maintained on the recursion stack, and the space needed for maintaining these
modified lengths grows with the recursion depth and certainly exceeds the
linear order. There is no apparent fix for such recursive algorithm to achieve
the linear space complexity.

\bigskip

In addition to the superlinear space complexity, all of the above strongly polynomial
algorithms have time complexity strictly growing with $M$. However, the
closeness of $M$ to $N$ could be \emph{beneficial} for the  computation of a
shortest $M$-link $1$-$N$ path. If $M=N-1$, then there is a single $\left(
N-1\right)  $-link $1$-$N$ path. If $M=N-2$, then there are $N-2$ $\left(
N-2\right)  $-link $1$-$N$ paths, each of which can be generated from the
unique $\left(  N-1\right)  $-link $1$-$N$ path by \emph{removing} an internal
node. Thus, a shortest $\left(  N-2\right)  $-link $1$-$N$ path can also be
computed in linear time and space. In fact, there is a \emph{symmetry} between
these two \textquotedblleft largest\textquotedblright\ cases and the two
\textquotedblleft smallest\textquotedblright cases. If $M=1$, then there is a
single $1$-link $1$-$N$ path. If $M=2$, then there are $N-2$ $2$-link $1$-$N$
paths, each of which can be generated by \emph{adding} an internal node. In
general, the number of $m$-link $1$-$N$ paths and the number of $\left(
N-m\right)  $-link $1$-$N$ paths are both equal to
\[
\binom{N-2}{m-1}=\binom{N-2}{N-m-1}.
\]
This symmetry sheds light on the room of improvement when $M$ is close to
$N$.  

\bigskip

In this paper, we present a contract-and-conquer algorithm for computing a
shortest $M$-link $1$-$N$ path \emph{iteratively} in $O\left(  \sqrt{NM\left(
N-M\right)  \log\left(  N-M\right)  }\right)  $ time and $O\left(  N\right)  $
space. Thus, it is the first $o\left(  NM\right)  $-time algorithm with
\emph{linear} space complexity. For the $O\left(  poly\left(  \log n\right)
\right)  $ and $N-O\left(  poly\left(  \log n\right)  \right)  $ regimes of
$M$, the contract-and-conquer algorithm has an $O\left(  N\cdot
poly\left(  \log n\right)  \right)  $ runtime, and hence partially answers the open
question in \cite{AST94} affirmatively. For all $M$, the contract-and-conquer
algorithm is superior to the algorithm AST \cite{AST94} in both time and space complexity. For $M$ around $N/2$, the
contract-and-conquer algorithm is inferior to Schieber's algorithm
\cite{Sch98} in time complexity in return for superior space complexity. The
contract-and-conquer algorithm follows the elementary decreasing-and-conquer
paradigm. It is conceptually simple and easy for implementation.

\bigskip

The following notations and terms are used in this paper. For two integers $a$
and $b$ with $a\leq b$, the closed integer interval $\left[  a:b\right]  $
denotes the set of integers $k$ with $a\leq k\leq b$; and $(a:b]$ represents
$\left[  a:b\right]  \setminus\left\{  a\right\}  $. For a positive integer
$n$, $\left[  n\right]  $ is a shorthand for $\left[  1:n\right]  $. A set
$\mathcal{L}\subseteq{\mathbb{Z}}^{2}$ is a \emph{lattice} \cite{Top11} if for any two
members $\left(  i_{1},j_{1}\right)  $ and $\left(  i_{2},j_{2}\right)  $ in
$\mathcal{L}$ with $i_{1} < i_{2}$ and $j_{1} > j_{2}$, both $\left(  i_{1}%
,j_{2}\right)  $ and $\left(  i_{2},j_{1}\right)  $ are also in $\mathcal{L}$. A
real-valued function $\rho$ on a lattice $\mathcal{L}$ is \emph{submodular} \cite{Top11}
if for any two members $\left(  i_{1},j_{1}\right)  $ and $\left(  i_{2}%
,j_{2}\right)  $ in $\mathcal{L}$ with $i_{1} < i_{2}$ and $j_{1} > j_{2}$,
\[
\rho\left(  i_{1},j_{1}\right)  +\rho\left(  i_{2},j_{2}\right)  \geq
\rho\left(  i_{1},j_{2}\right)  +\rho\left(  i_{2},j_{1}\right)  .
\]
A Monge DAG on an integer interval $\left[  s:N\right]  $ is a
complete DAG\ on $\left[  s:N\right]  $ with a submodular edge-length function
on the edge lattice $\left\{  \left(  i,j\right)  :s\leq i<j\leq N\right\}  $.
Throughout this paper, $1\leq s<N$ and $G_{s}$ is a Monge DAG on $\left[
s:N\right]  $ with edge-length function $c_{s}$. 

\bigskip

The remainder of this paper is organized as follows. Section \ref{s_PDP}
presents a parsimonious dynamic programming for computing the lengths of a
collection of shortest $m$-link paths. Section \ref{s_SPT} introduces the
notion of minimal and maximal shortest-path trees and characterizes their
depth properties. Section \ref{s_PSP} elaborates precisely on the parametric
search scheme for computing a shortest $m$-link path. Section \ref{s_Probe}
develops a key probe procedure with the hit-or-contract nature. Section
\ref{s_CC} describes the contract-and-conquer algorithm and analyzes its
complexity. Section \ref{s_conclude} concludes with discussion on possible improvements.

\bigskip

\section{A Parsimonious Dynamic Programming}

\label{s_PDP}

\bigskip

Suppose $1\leq s<N$ and $G_{s}$ is a Monge DAG on $\left[  s:N\right]  $ with edge-length function $c_{s}$. For each $n\in\left[  s+1:N\right]  $, there is
an $m$-link $s$-$n$ path if and only if $m\in\left[  n-s\right]  $. We use
$\mathcal{L}_{s}$ to denote the lattice
\[
\left\{  \left(  m,n\right)  :n\in\left[  s+1:N\right]  ,m\in\left[
n-s\right]  \right\}  .
\]
For each $\left(  m,n\right)  \in\mathcal{L}_{s}$, let $f_{s}\left(
m,n\right)  $ denote the minimum length of all $m$-link $s$-$n$ paths in
$G_{s}$, and $\mathcal{P}_{s}^{\ast}\left(  m,n\right)  $ denote the set of
all shortest $m$-link $s$-$n$ paths in $G_{s}$. The following computation task
is needed frequently in our later algorithm: Given $\left(  m,n\right)
\in\mathcal{L}_{s}$ with $m>1$ and $s+m<n\leq N$, compute $f_{s}\left(
m,j\right)  $ for \emph{all} $i\in\left[  s+m:n\right]  $ and $f_{s}\left(
m+1,j\right)  $ for \emph{all} $j\in\left[  s+m+1:n\right]  $. The outputs are
stored in two global arrays $f$ and $\overline{f}$ : $f_{s}\left(  m,j\right)
$ is stored at $f\left(  j\right)  $, and $f_{s}\left(  m+1,j\right)  $ is
stored at $\overline{f}$ $\left(  j\right)  $. In this section, we present a
parsimonious dynamic programming for this task which runs in $\Theta\left(
m\left(  n-s+1-m\right)  \right)  $ time and $\Theta\left(  n-s+1-m\right)  $
working space. 

\bigskip

All $f_{s}\left(  k,j\right)  $ for $\left(  k,j\right)  \in\mathcal{L}_{s}$
satisfy a variant of the Bellman-Ford recurrence:

\begin{itemize}
\item $f_{s}\left(  1,j\right)  =c_{s}\left(  s,j\right)  $ for each
$j\in\left[  s+1:N\right]  $; 

\item for each $\left(  k,j\right)  \in\mathcal{L}_{s}$ with $k>1$, \
\begin{equation}
f_{s}\left(  k,j\right)  =\min\limits_{i\in\left[  s+m-1:j-1\right]  }\left[
f_{s}\left(  k-1,i\right)  +c_{s}\left(  i,j\right)  \right]  .\label{eq_BF}%
\end{equation}

\end{itemize}

\begin{figure}
[ptbh]
\begin{center}
\includegraphics[
height=2.2009in,
width=2.7216in
]%
{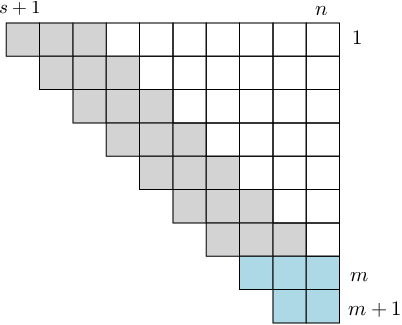}%
\caption{Minimal recurrence dependence.}%
\label{f_pbf}%
\end{center}
\end{figure}

\noindent The parsimonious dynamic programming exploits the \emph{minimal}
recurrence dependence as illustrated in Figure \ref{f_pbf}:

\begin{itemize}
\item The entries $f_{s}\left(  m,j\right)  $ for $j\in\left[  s+m:n\right]  $
depend exactly on the entries $f\left(  m-1,j\right)  $ for all $j\in\left[
s+m-1:n-1\right]  $.

\item In turn, the entries $f_{s}\left(  m-1,j\right)  $ for $j\in\left[
s+m-1:n-1\right]  $ further depend exactly on the entries $f_{s}\left(
m-2,j\right)  $ for all $j\in\left[  s+m-2:n-2\right]  $, and so on.
\end{itemize}

\noindent Recursively, the outputs depend exactly on the $\left(  m-1\right)
\left(  n-s+1-m\right)  $ entries $f_{s}\left(  k,j\right)  $ for all
$k\in\left[  1:m-1\right]  $ and $j\in\left[  s+k:n-m+k\right]  $.%

\bigskip

Accordingly, we get the explicit \textquotedblleft minimal\textquotedblright%
\ recurrence: 

\begin{itemize}
\item $f_{s}\left(  1,j\right)  =c_{s}\left(  s,j\right)  $ for each
$j\in\left[  s+1:n-m+1\right]  $.

\item For each $2\leq k\leq m$ and each $j\in\left[  s+k:n-m+k\right]  $,
\[
f_{s}\left(  k,j\right)  =\min_{i\in\left[  s+k-1:j-1\right]  }\left[
f_{s}\left(  k-1,i\right)  +c_{s}\left(  i,j\right)  \right]  .
\]

\item For each $j\in\left[  s+m+1:n\right]  $,
\[
f_{s}\left(  m+1,j\right)  =\min_{i\in\left[  s+m:j-1\right]  }\left[
f_{s}\left(  m,i\right)  +c_{s}\left(  i,j\right)  \right]  .
\]

\end{itemize}

\noindent For each $2\leq k\leq m$, the set
\[
\left\{  \left(  i,j\right)  :s+k-1\leq i<j\leq n-m+k\right\}
\]
is a sublattice of the edge lattice, and $f_{s}\left(  k-1,i\right)
+c_{s}\left(  i,j\right)  $ is submodular in $\left(  i,j\right)  $ on this
lattice; hence the SWAWK\ algorithm can be applied to compute $f_{s}\left(
k,j\right)  $ for all $j\in\left[  s+k:n-m+k\right]  $ in $\Theta\left(
n-s+1-m\right)  $ time and space. Similarly, the SWAWK\ algorithm can be
applied to compute $f_{s}\left(  m+1,j\right)  $ for all $j\in\left[
s+m+1:n\right]  $ in $\Theta\left(  n-s-m\right)  $ time and space. 

$\bigskip$

A subroutine \textbf{PBF}$\left(  s,m,n\right)  $ implementing the
parsimonious dynamic programming is outlined in Table \ref{tab_PBF}: 

\begin{itemize}
\item Initialize $f\left(  j\right)  =c_{s}\left(  s,j\right)  $ for each
$j\in\left[  s+1:n-m+1\right]  $.

\item For $k=2$ to $m$, first apply the SWAWK\ algorithm to compute
\[
\overline{f}\left(  j\right)  =\min_{i\in\left[  s+k-1:j-1\right]  }\left[
f\left(  i\right)  +c_{s}\left(  i,j\right)  \right]
\]
for all $j\in\left[  s+k:n-m+k\right]  $; and then overwrite $f\left(
j\right) $ with $\overline{f}\left(  j\right)$  for all $j\in\left[
s+k:n-m+k\right]  $.

\item Finally, apply the SWAWK\ algorithm to compute
\[
\overline{f}\left(  j\right)  =\min_{i\in\left[  s+m:j-1\right]  }\left[
f\left(  i\right)  +c_{s}\left(  i,j\right)  \right]
\]
for all $j\in\left[  s+m+1:n\right]  $.
\end{itemize}

\noindent Clearly, the subroutine \textbf{PBF}$\left(  s,m,n\right)  $ runs in
$\Theta\left(  m\left(  n-s+1-m\right)  \right)  $ time and needs additional
$O\left(  n-s+1-m\right)  $ working space by the SWAWK\ algorithm.%

\begin{table}[htbp] \centering
\begin{tabular}
[c]{|l|}\hline
\textbf{PBF}$\left(  s,m,n\right)  $:\\\hline\hline
for $j=s+1$ to $n-m+1$ do $f\left(  j\right)  \leftarrow c_{s}\left(
s,j\right)  $;\\
for $k=2$ to $m$ do \\
\quad for $j=s+k$ to $n-m+k$ do //SMAWK\\
\quad\quad$\overline{f}\left(  j\right)  \leftarrow\min_{i\in\left[
s+k-1:j-1\right]  }\left[  f\left(  i\right)  +c_{s}\left(  i,j\right)
\right]  $;\\
\quad for $j=s+k$ to $n-m+k$ do $f\left(  j\right)  \leftarrow\overline
{f}\left(  j\right)  $;\\
for $j=s+m+1$ to $n$ do //SMAWK\\
\quad$\overline{f}\left(  j\right)  \leftarrow\min_{i\in\left[
s+m:j-1\right]  }\left[  f\left(  i\right)  +c_{s}\left(  i,j\right)  \right]
$;\\\hline
\end{tabular}
\caption{Outline of \textbf{PBF}$(s,m,n)$.}\label{tab_PBF}%
\end{table}

\bigskip

\section{Minimal and Maximal Shortest-Path Trees}

\label{s_SPT}

\bigskip

Suppose $1\leq s<N$ and $G_{s}$ is a Monge DAG on $\left[
s:N\right]  $ with edge-length function $c_{s}$. A\ \emph{shortest-path tree}
(SPT) in $G_{s}$ is a directed tree on $\left[  s:N\right]  $ in which the
tree path from $s$ to each node $n$ is a shortest $s$-$n$ path in $G_{s}$. In
this section, we introduce the notion of minimal SPT and maximal SPT, and
assert that they are also respectively a shallowest SPT and a deepest SPT. 

\bigskip

For each $n\in\left[  s:N\right]  $, let $F_{s}\left(  n\right)  $ be the
minimum length of all $s$-$n$ paths in $G_{s}$. All $F_{s}\left(  n\right)  $
for $n\in\left[  s:N\right]  $ satisfy the recurrence: $F_{s}\left(  s\right)
=0$; and for each $n\in\left[  s+1:N\right]  $,%
\[
F_{s}\left(  n\right)  =\min_{i\in\left[  s:n-1\right]  }\left[  F_{s}\left(
i\right)  +c_{s}\left(  i,n\right)  \right]  .
\]

\noindent Denote $\Pi_{s}\left(  s\right)  :=\left\{  0\right\}  $; and for
each $n\in\left[  s+1:N\right]  $ denote%
\[
\Pi_{s}\left(  n\right)  :=\arg\min_{i\in\left[  s:n-1\right]  }\left[
F_{s}\left(  i\right)  +c_{s}\left(  i,n\right)  \right]  .
\]
Each SPT in $G_{s}$ is uniquely defined by a\ \emph{parent selection} $\pi$:
$\left[  s+1:N\right]  \rightarrow\left[  s:N-1\right]  $ satisfying that
$\pi\left(  n\right)  \in\Pi_{s}\left(  n\right)  $ for each $n\in\left[
s+1:N\right]  $, and vice versa. For each $n\in\left[  s:N\right]  $, let
$\pi_{s}^{\min}\left(  n\right)  $ (resp., $\pi_{s}^{\max}\left(  n\right)  $)
be the least (resp., greatest) member in $\Pi_{s}\left(  n\right)  $. Note
that $\pi_{s}^{\min}\left(  s\right)  =\pi_{s}^{\min}\left(  s\right)  =0$. As
$F_{s}\left(  i\right)  +c_{s}\left(  i,n\right)  $ is submodular in $\left(
i,n\right)  $ on the edge lattice of $G_{s}$, both $\pi_{s}^{\min}\left(
n\right)  $ and $\pi_{s}^{\max}\left(  n\right)  $ \emph{increase} with
$n\in\left[  s+1:N\right]  $. The \emph{minimal} (resp., \emph{maximal}) SPT
$T_{s}^{\min}$ (resp., $T_{s}^{\max}$) in $G_{s}$ is the SPT in $G_{s}$
defined by $\pi_{s}^{\min}$ (resp., $\pi_{s}^{\max}$). 

\bigskip

For each $n\in\left[  s:N\right]  $, let $D_{s}\left(  n\right)  $ be the set
of numbers of links in all shortest $s$-$n$ paths in $G_{s}$, and $d_{s}%
^{\min}\left(  n\right)  $ (respectively, $d_{s}^{\max}\left(  n\right)  $) be
the least (respectively, greatest) member in $D_{s}\left(  n\right)  $.
Clearly, $D_{s}\left(  s\right)  =\left\{  0\right\}  $. For each $n\in\left[
s+1:N\right]  $, each shortest $s$-$n$ path in $G_{s}$ with $m\in D_{s}\left(
n\right)  $ links must belong to $\mathcal{P}_{s}^{\ast}\left(  m,n\right)  $
and have length $f_{s}\left(  m,n\right)  $; hence%
\begin{align}
F_{s}\left(  n\right)   &  =\min_{\left(  m,n\right)  \in\mathcal{L}_{s}}%
f_{s}\left(  m,n\right)  =\min_{m\in\left[  n-s\right]  }f_{s}\left(
m,n\right)  ,\nonumber\\
D_{s}\left(  n\right)   &  =\arg\min_{m\in\left[  n-s\right]  }f_{s}\left(
m,n\right)  .\label{eq_SPinterval}%
\end{align}
Thus, the collection of shortest $s$-$n$ paths in $G_{s}$ is exactly the union
of $\mathcal{P}_{s}^{\ast}\left(  m,n\right)  $ over all $m\in D_{s}\left(
n\right)  $. The theorem below asserts that $T_{s}^{\min}$ (resp, $T_{s}%
^{\max}$) is also a \emph{shallowest} (resp., \emph{deepest}) SPT.

\bigskip

\begin{theorem}
\label{t_depth} The depth of each $n\in\left[  s:N\right]  $ in $T_{s}^{\min}$
(resp., $T_{s}^{\max}$) is exactly $d_{s}^{\min}\left(  n\right)  $ (resp.,
$d_{s}^{\max}\left(  n\right)  $). In addition, all nodes with the same depth
in $T_{s}^{\min}$ (resp, $T_{s}^{\max}$) are consecutive.
\end{theorem}

\bigskip

To prove the above theorem, we describe a \emph{path swapping} operation which
is more direct and explicit than the variants in \cite{AST94,BLP92,Sch98}.
Consider two paths
\begin{align*}
P  & =\left(  u_{0},u_{1},\cdots,u_{m_{1}}\right)  ,\\
Q  & =\left(  v_{0},v_{1},\cdots,v_{m_{2}}\right)
\end{align*}
in $G_{s}$ with $1\leq m_{1}\leq m_{2}$ and $u_{0}\leq v_{0}<v_{m_{2}}\leq
u_{m_{1}}$ (see Figure \ref{f_pathswap2}). For any $m\in\left[  m_{1}%
:m_{2}\right]  $, we construct an $m$-link $v_{0}$-$u_{m_{1}}$ path
$Q\oplus_{m}P$ and an $\left(  m_{1}+m_{2}-m\right)  $-link $u_{0}$-$v_{m_{2}%
}$ path $Q\ominus_{m}P$ in two steps:%

\begin{figure}
[ptbh]
\begin{center}
\includegraphics[
height=1.8965in,
width=4.2938in
]%
{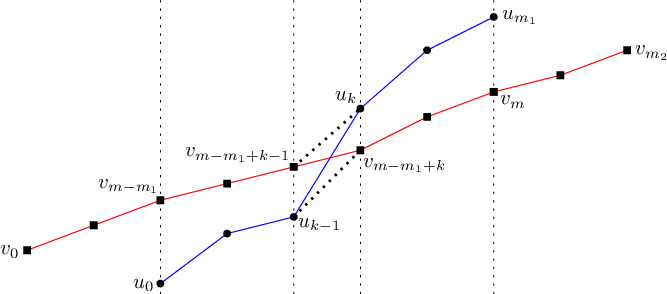}%
\caption{Path swapping.}%
\label{f_pathswap2}%
\end{center}
\end{figure}

\begin{itemize}
\item \textbf{Step 1}:\ Compute the \emph{least} $k\in\left[  m_{1}\right]  $
with $u_{k}\geq v_{m-m_{1}+k}$. Such $k$ does exist as $u_{m_{1}}\geq
v_{m_{2}}\geq v_{m}=v_{m-m_{1}+m_{1}}$. Then $u_{k-1}\leq v_{m-m_{1}+k-1}%
$:\ if $k=1$ then%
\[
u_{k-1}=u_{0}\leq v_{0}\leq v_{m-m_{1}}=v_{m-m_{1}+k-1};
\]
otherwise $u_{k-1}<v_{m-m_{1}+k-1}$ by the least choice of $k$. Hence,%
\[
u_{k-1}\leq v_{m-m_{1}+k-1}<v_{m-m_{1}+k}\leq u_{k}.
\]

\item \textbf{Step 2}: Replace the edges $\left(  u_{k-1},u_{k}\right)  $ in
$P$ and $\left(  v_{m-m_{1}+k-1},v_{m-m_{1}+k}\right)  $ in $Q$ respectively
with the edges $\left(  u_{k-1},v_{m-m_{1}+k}\right)  $ and $\left(
v_{m-m_{1}+k-1},u_{k}\right)  $ to get%
\begin{align*}
Q\oplus_{m}P  & :=\left(  v_{0},\cdots,v_{m-m_{1}+k-1},u_{k},\cdots,u_{m_{1}%
}\right)  ,\\
Q\ominus_{m}P  & :=\left(  u_{0},\cdots,u_{k-1},v_{m-m_{1}+k},\cdots,v_{m_{2}%
}\right)  .
\end{align*}

\end{itemize}

\noindent By the submodularity of $c_{s}$, the two new edges have no larger
sum-length than the two old edges. Thus,
\begin{equation}
c_{s}\left(  P\right)  +c_{s}\left(  Q\right)  \geq c_{s}\left(  Q\oplus
_{m}P\right)  +c_{s}\left(  Q\ominus_{m}P\right)  .\label{eq_swap}%
\end{equation}

\bigskip

The swapping of two shortest $s$-$n$ paths with $d_{s}^{\min}\left(  n\right)
$ and $d_{s}^{\max}\left(  n\right)  $ links respectively has the following
direct consequence.

\bigskip

\begin{lemma}
\label{l_SPexchange} Suppose $n\in\left[  s+1:N\right]  $, and $P$ (resp.,
$Q$) is a shortest $s$-$n$ path with $d_{s}^{\min}\left(  n\right)  $ (resp.
$d_{s}^{\max}\left(  n\right)  $) links. Then for any $m\in\left[  d_{s}%
^{\min}\left(  n\right)  :d_{s}^{\max}\left(  n\right)  \right]  $,
$Q\oplus_{m}P$ is a shortest $s$-$n$ path with $m$ links. Hence, $D_{s}\left(
n\right)  =\left[  d_{s}^{\min}\left(  n\right)  :d_{s}^{\max}\left(
n\right)  \right]  $. 
\end{lemma}

\bigskip

For any $N\geq n_{1}\geq n_{2}>s$ and $1\leq m_{1}\leq m_{2}\leq n_{2}-s$, the
swapping of a path in $\mathcal{P}_{s}^{\ast}\left(  m_{1},n_{1}\right)  $ and
a path in $\mathcal{P}_{s}^{\ast}\left(  m_{2},n_{2}\right)  $ implies
immediately the lemma below which is implicit in \cite{AST94,BLP92,Sch98}.

\bigskip

\begin{lemma}
\label{l_bimontone} Suppose $N\geq n_{1}\geq n_{2}>s$ and $1\leq m_{1}\leq
m_{2}\leq n_{2}-s$. Then for any $m\in\left[  m_{1}:m_{2}\right]  $,
\[
f_{s}\left(  m_{1},n_{1}\right)  +f_{s}\left(  m_{2},n_{2}\right)  \geq
f_{s}\left(  m,n_{1}\right)  +f_{s}\left(  m_{1}+m\,_{2}-m,n_{2}\right)  .
\]
In particular, $f_{s}\left(  m,n\right)  $ is a submodular function on
$\mathcal{L}_{s}$.
\end{lemma}

\bigskip

The submodularity of $f_{s}\left(  m,n\right)  $ on $\mathcal{L}_{s}$ and the
minimizer representation of $D_{s}\left(  n\right)  $ in equation
(\ref{eq_SPinterval}) imply the monotonicity of $d_{s}^{\min}\left(  n\right)
$ and $d_{s}^{\max}\left(  n\right)  $ with $n\in\left[  s+1:N\right]  $. As
$d_{s}^{\min}\left(  0\right)  =d_{s}^{\max}\left(  n\right)  =0$, we get the
following lemma. 

\bigskip

\begin{lemma}
\label{l_depthmonotone} Both $d_{s}^{\min}\left(  n\right)  $ and $d_{s}%
^{\max}\left(  n\right)  $ increase with $n\in\left[  s:N\right]  $.
\end{lemma}

\bigskip

Now are ready to prove Theorem \ref{t_depth}. Consider any $n\in\left[
s+1:N\right]  $. By Lemma \ref{l_depthmonotone},
\begin{align*}
d_{s}^{\min}\left(  n\right)   &  =1+\min_{i\in\Pi_{s}\left(  n\right)  }%
d_{s}^{\min}\left(  i\right)  =1+d_{s}^{\min}\left(  \pi_{s}^{\min}\left(
n\right)  \right)  ,\\
d_{s}^{\max}\left(  n\right)   &  =1+\max_{i\in\Pi_{s}\left(  n\right)  }%
d_{s}^{\max}\left(  i\right)  =1+d_{s}^{\max}\left(  \pi_{s}^{\max}\left(
n\right)  \right)  .
\end{align*}
Thus, the first part of Theorem \ref{t_depth} holds. The second part of
Theorem \ref{t_depth} then follows from Lemma \ref{l_depthmonotone}.

\bigskip

All algorithms in \cite{GP90,Kla89,LS91,Wil88} can compute $F_{s}$ in linear
time and space. Given $F_{s}$, $\pi_{s}^{\min}$ (resp, $\pi_{s}^{\max}$) can
be computed in linear time and space by a single application of the
SWAWK\ algorithm. Given $\pi_{s}^{\min}$ (resp., $\pi_{s}^{\max}$) ,
$d_{s}^{\min}$ (resp., $d_{s}^{\max}$) can be computed in linear space and
time by a tree traversal. If we only need the depth of $N$ and/or the tree
path from $s$ to $N$ in $T_{s}^{\min}$ (resp, $T_{s}^{\max}$), then a
backtrack from $N$ to $s$ along $\pi_{s}^{\min}$ (or $\pi_{s}^{\max}$) is sufficient.

$\bigskip$

In \cite{AST94,Sch98}, a tie-breaking modification to the Wilber's algorithm
\cite{Wil88} or Klawe's algorithm \cite{Kla89} was proposed for computing
a shortest $s$-$N$ path with the fewest (resp., most) links: Whenever the
lengths of two paths are compared and found to be equal, the path with fewer
(resp., more) links is considered lighter. While ambiguity remains when both
the lengths and the number of links of two paths are compared and found to be
equal, such further tie-breaking apparently still relies on the original
lexicographic rule. The argument for the (non-obvious) monotonicity of the
modified parent selection, which is essential to the application of the
SWAWK\ algorithm, was missing in \cite{AST94,Sch98}. In contrast, the
construction of $T_{s}^{\min}$ (resp, $T_{s}^{\max}$) eliminates the need for
tie-breaking by the number of links. 

\bigskip

\section{Parametric Shortest Paths}

\label{s_PSP}

\bigskip

For each parameter $\lambda\in\mathbb{R}$, let $G_{s}\left(  \lambda\right)  $
denote the DAG\ obtained from $G_{s}$ by \emph{subtracting} $\lambda$ from the
length of each edge. Then, $G_{s}\left(  \lambda\right)  $ is also submodular.
For any $\left(  m,n\right)  \in\mathcal{L}_{s}$, the collection of shortest
$m$-link $s$-$n$ paths in $G_{s}\left(  \lambda\right)  $ is still
$\mathcal{P}_{s}^{\ast}\left(  m,n\right)  $, and the length of each path in
$\mathcal{P}_{s}^{\ast}\left(  m,n\right)  $ in $G_{s}\left(  \lambda\right)
$ is $f_{s}\left(  m,n\right)  -m\lambda$. Thus, for the purpose of seeking a
path in $\mathcal{P}_{s}^{\ast}\left(  M,N\right)  $, we may seek a shortest
$M$-link $s$-$N$ path in $G_{s}\left(  \lambda\right)  $ for \emph{any}
$\lambda\in\mathbb{R}$. The general idea of the parametric search scheme
\cite{AST94,BLP92} is to find some parameter $\lambda$ such that $G_{s}\left(
\lambda\right)  $ has a shortest $s$-$N$ path with $M$ links. For any such
$\lambda$, each shortest $s$-$N$ path in $G_{s}\left(  \lambda\right)  $ with
$M$ links is a path in $\mathcal{P}_{s}^{\ast}\left(  M,N\right)  $, and one
such path can be computed in linear time and space. In this section, we
provide a concise and precise elaboration on the parametric search scheme
\cite{AST94,BLP92}.

\bigskip

For each $n\in\left[  s:N\right]  $, let $F_{s}\left(  \lambda,n\right)  $ be
the minimum length of all $s$-$n$ paths in $G_{s}\left(  \lambda\right)  $,
and $D_{s}\left(  \lambda,n\right)  :=\left[  d_{s}^{\min}\left(
\lambda,n\right)  :d_{s}^{\max}\left(  \lambda,n\right)  \right]  $ be the set
of numbers of links in all shortest $s$-$n$ paths in $G_{s}\left(
\lambda\right)  $. Clearly, $F_{s}\left(  \lambda,s\right)  =0$ and
$D_{s}\left(  \lambda,s\right)  =\left\{  0\right\}  $. For each $n\in\left[
s+1:N\right]  $,%
\begin{align*}
F_{s}\left(  \lambda,n\right)    & =\min_{m\in\left[  n-s\right]  }\left[
f_{s}\left(  m,n\right)  -m\lambda\right]  ,\\
D_{s}\left(  \lambda,n\right)    & =\arg\min_{m\in\left[  n-s\right]  }\left[
f_{s}\left(  m,n\right)  -m\lambda\right]  .
\end{align*}
Thus, the collection of shortest $s$-$n$ paths in $G_{s}\left(  \lambda
\right)  $ is exactly the union of $\mathcal{P}_{s}^{\ast}\left(  m,n\right)
$ over all $m\in D_{s}\left(  \lambda,n\right)  $. Let $T_{s}^{\min}\left(
\lambda\right)  $ (resp., $T_{s}^{\max}\left(  \ \lambda\right)  $) denote the
minimal (resp., maximal) SPT of $G_{s}\left(  \lambda\right)  $. Then for each
\thinspace$n\in\left[  s:N\right]  $, $d_{s}^{\min}\left(  \lambda,n\right)  $
(resp., $d_{s}^{\max}\left(  \lambda,n\right)  $) is the depth of $n$ in
$T_{s}^{\min}\left(  \lambda\right)  $ (resp., $T_{s}^{\max}\left(
\lambda\right)  $) by Theorem \ref{t_depth}. 

\bigskip

Suppose $M\in D_{s}\left(  \lambda,N\right)  $.  Then each path in
$\mathcal{P}_{s}^{\ast}\left(  M,N\right)  $ is a shortest $s$-$N$ path in
$G_{s}\left(  \lambda\right)  $. A path in $\mathcal{P}_{s}^{\ast}\left(
M,N\right)  $ can be computed in linear time and space as follows:

\begin{itemize}
\item Compute $T_{s}^{\min}\left(  \lambda\right)  $ and $T_{s}^{\max}\left(
\ \lambda\right)  $ in $G_{s}\left(  \lambda\right)  $, and let $P$ and $Q$ be
the $s$-$N$ paths in $T_{s}^{\min}\left(  \lambda\right)  $ and $T_{s}^{\max
}\left(  \lambda\right)  $ respectively.

\item Construct $Q\oplus_{M}P$. By Lemma \ref{l_SPexchange}, $Q\oplus_{M}P$ is
a shortest $s$-$N$ path in $G\left(  \lambda\right)  $ with $M$ links; hence
$Q\oplus_{M}P\in\mathcal{P}_{s}^{\ast}\left(  M,N\right)  $. 
\end{itemize}

\noindent Thus, computing a path in $\mathcal{P}_{s}^{\ast}\left(  M,N\right)
$ is reducible in \emph{linear} time and space to finding a parameter
$\lambda$ with $M\in D_{s}\left(  \lambda,N\right)  $.

\bigskip

In the sequel, we characterize the range of $\lambda$ such that $m\in
D_{s}\left(  \lambda,n\right)  $ for a given $\left(  m,n\right)
\in\mathcal{L}_{s}$. For any $n\in\left[  s+1:N\right]  $, let
\[
f_{s}\left(  0,n\right)  =f_{s}\left(  n-s+1,n\right)  =+\infty
\]
as no $s$-$n$ path in $G_{s}$ has $0$ or $n-s+1$ links; and let
\[
\delta_{s}\left(  m,n\right)  :=f_{s}\left(  m+1,n\right)  -f_{s}\left(
m,n\right)
\]
for each $0\leq m\leq n-s$. Note that $\delta_{s}\left(  m,n\right)  $ is
finite for each $m\in\left[  n-s-1\right]  $, $\delta_{s}\left(  0,n\right)
=-\infty$, and $\delta_{s}\left(  n-s,n\right)  =+\infty$. By Lemma
\ref{l_bimontone}, $\delta_{s}\left(  m,n\right)  $ increases with $m$ and
decreases with $n$. The theorem below gives an equivalent condition for
$d_{s}^{\min}\left(  \lambda,n\right)  \leq m$ and an equivalent condition for
$d_{s}^{\max}\left(  \lambda,n\right)  \geq m$. 

\bigskip

\begin{theorem}
\label{t_window} For any $\left(  m,n\right)  \in\mathcal{L}_{s}$ and
$\lambda\in\mathbb{R}$,

\begin{itemize}
\item $d_{s}^{\min}\left(  \lambda,n\right)  \leq m$ if and only if
$\lambda\leq\delta_{s}\left(  m,n\right)  $;

\item $d_{s}^{\max}\left(  \lambda,n\right)  \geq m$ if and only if
$\lambda\geq\delta_{s}\left(  m-1,n\right)  $.
\end{itemize}
\end{theorem}

\bigskip

\begin{proof}
We first show that
\begin{equation}
\delta_{s}\left(  d_{s}^{\min}\left(  \lambda,n\right)  -1,n\right)
<\lambda<\delta_{s}\left(  d_{s}^{\max}\left(  \lambda,n\right)  ,n\right)
.\label{eq_strict}%
\end{equation}
For $j=d_{s}^{\min}\left(  \lambda,n\right)  $, $j\in D_{s}\left(
\lambda,n\right)  $ and $j-1\notin D_{s}\left(  \lambda,n\right)  $; hence
\[
\delta_{s}\left(  j-1,n\right)  -\lambda=F_{s}\left(  \lambda,n\right)
-\left[  f_{s}\left(  j-1,n\right)  +\left(  j-1\right)  \lambda\right]  <0.
\]
For $j=d_{s}^{\max}\left(  \lambda\right)  $, $j\in D_{s}\left(
\lambda,n\right)  $ and $j+1\notin D_{s}\left(  \lambda,n\right)  $; hence
\[
\delta_{s}\left(  j,n\right)  -\lambda=\left[  f_{s}\left(  j+1,n\right)
+\left(  j+1\right)  \lambda\right]  -F_{s}\left(  \lambda,n\right)  >0.
\]
Thus, the two strict inequalities in equation (\ref{eq_strict}) hold. 

Now, we show that
\begin{equation}
\delta_{s}\left(  d_{s}^{\max}\left(  \lambda,n\right)  -1,n\right)
\leq\lambda\leq\delta_{s}\left(  d_{s}^{\min}\left(  \lambda,n\right)
,n\right)  .\label{eq_nonstrict}%
\end{equation}
If $d_{s}^{\min}\left(  \lambda,n\right)  =d_{s}^{\max}\left(  \lambda
,n\right)  $, the inequalities hold strictly by equation (\ref{eq_strict}).
Suppose $d_{s}^{\min}\left(  \lambda,n\right)  <d_{s}^{\max}\left(
\lambda,n\right)  $. Then $f_{s}\left(  j,n\right)  +j\lambda=F_{s}\left(
\lambda,n\right)  $ for $j=d_{s}^{\min}\left(  \lambda,n\right)  $,
$d_{s}^{\min}\left(  \lambda,n\right)  +1$, $d_{s}^{\max}\left(
\lambda,n\right)  -1$, and $d_{s}^{\max}\left(  \lambda,n\right)  $. Thus, the
two inequalities in equation (\ref{eq_nonstrict}) hold with equality. 

Next, we prove the first part of the theorem. If $d_{s}^{\min}\left(
\lambda,n\right)  \leq m$, then by equation (\ref{eq_nonstrict})
\[
\lambda\leq\delta_{s}\left(  d_{s}^{\min}\left(  \lambda,n\right)  ,n\right)
\leq\delta_{s}\left(  m,n\right)  .
\]
If $d_{s}^{\min}\left(  \lambda,n\right)  >m$, then by equation
(\ref{eq_strict})
\[
\lambda>\delta_{s}\left(  d_{s}^{\min}\left(  \lambda,n\right)  -1,n\right)
\geq\delta_{s}\left(  m,n\right)  .
\]

Finally, we prove the second part of the theorem. If $d_{s}^{\max}\left(
\lambda,n\right)  \geq m$, then by equation (\ref{eq_nonstrict})
\[
\lambda\geq\delta_{s}\left(  d_{s}^{\max}\left(  \lambda,n\right)
-1,n\right)  \geq\delta_{s}\left(  m-1,n\right)  .
\]
If $d_{s}^{\max}\left(  \lambda,n\right)  <m$, then by equation
(\ref{eq_strict})
\[
\lambda<\delta_{s}\left(  d_{s}^{\max}\left(  \lambda,n\right)  ,n\right)
\leq\delta_{s}\left(  m-1,n\right)  .
\]

Therefore, the theorem holds. 
\end{proof}

\bigskip

By Theorem \ref{t_window}, for any $\left(  m,n\right)  \in\mathcal{L}_{s}$
and $\lambda\in\mathbb{R}$, $m\in D_{s}\left(  \lambda,n\right)  $ if and only
if $\delta_{s}\left(  m-1,n\right)  \leq\lambda\leq\delta_{s}\left(
m,n\right)  $. With a slight notational abuse, each closed extended-real
interval \emph{excludes} its \emph{infinite} endpoints. For any $\left(
m,n\right)  \in\mathcal{L}_{s}$, denote
\[
\Lambda_{s}\left(  m,n\right)  :=\left[  \delta_{s}\left(  m-1,n\right)
,\delta_{s}\left(  m,n\right)  \right]  .
\]
Then $\Lambda_{s}\left(  m,n\right)  $ is \emph{exactly} the range of
$\lambda$ with $m\in D_{s}\left(  \lambda,n\right)  $.

\bigskip

In the remaining of this section, we derive monotonic properties of
$D_{s}\left(  \lambda,n\right)  $ and $\Lambda_{s}\left(  m,n\right)  $. By
Lemma \ref{l_depthmonotone}, both endpoints of $D_{s}\left(  \lambda,n\right)
$ increase with $n\in\left[  s,N\right]  $; they also increase with
$\lambda\in\mathbb{R}$ implied by the lemma below. 

\bigskip

\begin{lemma}
\label{l_monotone_lambda} For any $n\in\left[  s,N\right]  $ and $\lambda
_{1}<\lambda_{2}$, $d_{s}^{\max}\left(  \lambda_{1},n\right)  \leq d_{s}%
^{\min}\left(  \lambda_{2},n\right)  .$
\end{lemma}

\bigskip

\begin{proof}
As $D_{s}\left(  \lambda,s\right)  =\left\{  0\right\}  $, the lemma holds
trivially for $n=s$; henceforth we assume $n>s$. Denote $m_{1}:=d_{s}^{\max
}\left(  \lambda_{1},n\right)  $ and $m_{2}:=d_{s}^{\min}\left(  \lambda
_{2},n\right)  $. Then,
\begin{align*}
f_{s}\left(  m_{1},n\right)  -m_{1}\lambda_{1} &  =F_{s}\left(  \lambda
_{1},n\right)  \leq f_{s}\left(  m_{2},n\right)  -m_{2}\lambda_{1},\\
f_{s}\left(  m_{2},n\right)  -m_{2}\lambda_{2} &  =F_{s}\left(  \lambda
_{2},n\right)  \leq f_{s}\left(  m_{1},n\right)  -m_{1}\lambda_{2}.
\end{align*}
Summing up the above two inequalities yields
\[
m_{1}\lambda_{1}+m_{2}\lambda_{2}\geq m_{2}\lambda_{1}+m_{1}\lambda_{2},
\]
and hence $m_{1}\left(  \lambda_{2}-\lambda_{1}\right)  \leq m_{2}\left(
\lambda_{2}-\lambda_{1}\right)  $. Thus, $m_{1}\leq m_{2}$.
\end{proof}

\bigskip

By Lemma \ref{l_bimontone}, both endpoints of $\Lambda_{s}\left(  m,n\right)
$ increase with $m\in\left[  n-s\right]  $ and decrease with $n\in\left[
s+m+1,N\right]  $. In addition, the following \textquotedblleft
vertical\textquotedblright\ downward monotonicity holds.

\bigskip

\begin{lemma}
\label{l_vertical} For any $\left(  m,n\right)  \in\mathcal{L}_{s}$ and any
path $\left(  v_{0},\cdots,v_{m}\right)  $ in $\mathcal{P}_{s}^{\ast}\left(
m,n\right)  $,
\[
\Lambda_{s}\left(  1,v_{1}\right)  \supseteq\Lambda_{s}\left(  2,v_{2}\right)
\supseteq\cdots\supseteq\Lambda_{s}\left(  m,v_{m}\right)  .
\]

\end{lemma}

\bigskip

\begin{proof}
We show that $\Lambda_{s}\left(  i,v_{i}\right)  \supseteq\Lambda_{s}\left(
i+1,v_{i+1}\right)  $ for any $1\leq i<m$. Consider any $\lambda\in\Lambda
_{s}\left(  i+1,v_{i+1}\right)  $. Then $i+1\in D_{s}\left(  \lambda
,v_{i+1}\right)  $; hence the subpath $\left(  v_{0},\cdots,v_{i}%
,v_{i+1}\right)  $ is a shortest $s$-$v_{i+1}$ path in $G_{s}\left(
\lambda\right)  $. Accordingly, the subpath $\left(  v_{0},\cdots
,v_{i}\right)  $ is a shortest $s$-$v_{i}$ path in $G_{s}\left(
\lambda\right)  $. So, $i\in D_{s}\left(  \lambda,v_{i}\right)  $; hence
$\lambda\in\Lambda_{s}\left(  i,v_{i}\right)  $. Thus, $\Lambda_{s}\left(
i+1,v_{i+1}\right)  \subseteq\Lambda_{s}\left(  i,v_{i}\right)  $.
\end{proof}

\bigskip

\begin{corollary}
\label{c_probebound}For any $1\leq m\leq M\leq N-s-1$ and any path $\left(
v_{0},\cdots,v_{M+1}\right)  $ in $\mathcal{P}_{s}^{\ast}\left(  M+1,N\right)
$, $v_{m+1}\in\left[  s+m+1:N-M+m\right]  $ and $\delta_{s}\left(
m,v_{m+1}\right)  \leq\delta_{s}\left(  M,N\right)  $.
\end{corollary}

\bigskip

\section{The $m$-Probe}

\label{s_Probe}

\bigskip

Suppose $G_{s}$ is a Monge DAG\ on $\left[  s:N\right]  $ with edge-length function $c_{s}$, and $4\leq M\leq N-s-1$. Consider a parameter $m\in\left[
2:M-2\right]  $. The $m$\emph{-probe} of $G_{s}$ is the \emph{least}
$r\in\left[  s+m+1:N-M+m\right]  $ satisfying that $\delta_{s}\left(
m,r\right)  \leq\delta_{s}\left(  M,N\right)  $. By Corollary
\ref{c_probebound}, such $r$ is well-defined. The $m$-probe $r$ plays a
pivotal role in seeking a member of $\Lambda_{s}\left(  M,N\right)  $. 

\bigskip

For each $j\in\left[  r:N\right]  $, let
\[
h\left(  j\right)  :=\min_{s+m\leq i<r}\left[  f_{s}\left(  m,i\right)
+c_{s}\left(  i,j\right)  \right]  .
\]
Note that $h\left(  j\right)  $ is exactly the minimum length of all $\left(
m+1\right)  $-link $s$-$j$ paths with the penultimate node among $\left[
s+m:r-1\right]  $. The \emph{contraction} $G_{r-1}$ of $G_{s}$ is a DAG\ on
$\left[  r-1:N\right]  $ obtained from $G_{s}$ by 

\begin{itemize}
\item deleting all nodes in $\left[  s:r-2\right]  $, and 

\item reassigning a length $h\left(  j\right)  $ to the edge $\left(
r-1,j\right)  $ for each $j\in\left[  r:N\right]  $.
\end{itemize}

\noindent Let $c_{r-1}$ be the edge length function of $G_{r-1}$. Since $r\leq
N-M+m$,
\[
2\leq M-m\leq N-r=N-\left(  r-1\right)  -1.
\]
The theorem below is the cornerstone to the design of our contract-and-conquer algorithm.

\bigskip\ 

\begin{theorem}
\label{t_probe} $G_{r-1}$ is Monge. If $\delta_{s}\left(  m,r\right)
\notin\Lambda_{s}\left(  M,N\right)  $, then $\Lambda_{r-1}\left(
M-m,N\right)  $ in $G_{r-1}$ coincides with $\Lambda_{s}\left(  M,N\right)  $
in $G_{s}$.
\end{theorem}

\bigskip

\begin{proof}
In order to prove $G_{r-1}$ is Monge, it suffices to show that the quadruple
inequality
\[
c_{r-1}\left(  r-1,j+1\right)  -c_{r-1}\left(  r-1,j\right)  \geq
c_{r-1}\left(  r,j+1\right)  -c_{r-1}\left(  r,j\right)
\]
holds for any $r<j<N$. Let $i\in\lbrack s+m:r-1]$ be such that
\[
c_{r-1}\left(  r-1,j+1\right)  =f_{s}\left(  m,i\right)  +c_{s}\left(
i,j+1\right)  .
\]
Then
\[
c_{r-1}\left(  r-1,j\right)  \leq f_{s}\left(  m,i\right)  +c_{s}\left(
i,j\right)  .
\]
Hence,
\begin{align*}
&  c_{r-1}\left(  r-1,j+1\right)  -c_{r-1}\left(  r-1,j\right)  \\
&  \geq c_{s}\left(  i,j+1\right)  -c_{s}\left(  i,j\right)  \\
&  \geq c_{s}\left(  r,j+1\right)  -c_{s}\left(  r,j\right)  \\
&  =c_{r-1}\left(  r,j+1\right)  -c_{r-1}\left(  r,j\right)  .
\end{align*}
where the second inequality is due to the submodularity of $c_{s}$.

Suppose $\delta_{s}\left(  m,r\right)  \notin\Lambda_{s}\left(  M,N\right)  $.
Then,
\[
\delta_{s}\left(  m,r\right)  <\delta_{s}\left(  M-1,N\right)  .
\]
Consider any $M^{\prime}\in\left[  M-1:M+1\right]  $. Then, $M^{\prime}%
-m\in\left[  M-m-1:M-m+1\right]  $. We prove that for any shortest $\left(
M^{\prime}-m\right)  $-link $\left(  r-1\right)  $-$N$ path $P=\left(
u_{0},u_{1},\cdots,u_{M^{\prime}-m}\right)  $ in $G_{r-1}$ and any shortest
$M^{\prime}$-link $s$-$N$ path $Q=\left(  v_{0},v_{1},\cdots,v_{M^{\prime}%
}\right)  $ in $G_{s}$, $c_{r-1}\left(  P\right)  =c_{s}\left(  Q\right)  $.
Consequently, $\Lambda_{r-1}\left(  M-m,N\right)  $ in $G_{r-1}$ is identical
to $\Lambda_{s}\left(  M,N\right)  $ in $G_{s}$.

We first claim that $v_{m}<r\leq v_{m+1}$. By Lemma \ref{l_vertical},
\[
\delta_{s}\left(  m,v_{m+1}\right)  \leq\delta_{s}\left(  M^{\prime
}-1,N\right)  \leq\delta_{s}\left(  M,N\right)  ;
\]
hence $r\leq v_{m+1}$ by the least choice of $r$. Note that $s+m\leq
v_{m}<v_{m+1}$. The inequality $v_{m}<r$ holds trivially if $v_{m}=s+m$.
Suppose $v_{m}\geq s+m+1$. Then by Lemma \ref{l_vertical} and the two
conditions $M^{\prime}\geq M-1$ and $\delta_{s}\left(  m,r\right)  <\delta
_{s}\left(  M-1,N\right)  $, we have
\[
\delta_{s}\left(  m,v_{m}\right)  \geq\delta_{s}\left(  M^{\prime},N\right)
\geq\delta_{s}\left(  M-1,N\right)  >\delta_{s}\left(  m,r\right)  ;
\]
hence $v_{m}<r$. Thus, the claim holds.

Now, we show that $c_{s}\left(  Q\right)  \geq c_{r-1}\left(  P\right)  $. As
$Q$ is a shortest $M^{\prime}$-link $s$-$N$ path in $G_{s}$, the subpath
$\left(  v_{0},\cdots,v_{m+1}\right)  $ is a shortest $\left(  m+1\right)
$-link $s$-$v_{m+1}$ path in $G_{s}$. The previous claim yields that
\[
c_{r-1}\left(  r-1,v_{m+1}\right)  =c_{s}\left(  v_{0},\cdots,v_{m+1}\right)
.
\]
Let $Q^{\prime}$ be the $\left(  M^{\prime}-m\right)  $-link $\left(
r-1\right)  $-$N$ path in $G_{r-1}$ obtained from $Q$ by replacing the first
$m+1$ nodes $v_{0},\cdots,v_{m}$ with the node $r-1$. Then,
\[
c_{s}\left(  Q\right)  =c_{r-1}\left(  Q^{\prime}\right)  \geq c_{r-1}\left(
P\right)  .
\]

Next, we show that $c_{s}\left(  Q\right)  \leq c_{r-1}\left(  P\right)  $.
Let $P^{\prime}$ be the concatenation of a shortest $\left(  m+1\right)
$-link $s$-$u_{1}$ path in $G_{s}$ and the subpath $\left(  u_{1}%
,\cdots,u_{M^{\prime}-m}\right)  $ of $P$. By the definition of $c_{r-1}$,
$c_{r-1}\left(  P\right)  =c_{s}\left(  P^{\prime}\right)  $. As $c_{s}\left(
P^{\prime}\right)  \geq c_{s}\left(  Q\right)  $, we have $c_{r-1}\left(
P\right)  \geq c_{s}\left(  Q\right)  $.

Thus, the theorem follows.
\end{proof}

\bigskip

In the sequel, we present a procedure \textbf{Probe}$\left(  s,m\right)  $ on
$G_{s}$ which 

\begin{itemize}
\item either hits and returns a member $\delta_{s}\left(  m,n\right)  $ of
$\Lambda_{s}\left(  M,N\right)  $ if there is any,

\item or finds $r$, contracts $G_{s}$ into $G_{r-1}$, and reduces $M$ by $m$.
\end{itemize}

\noindent In the latter case, the new lengths of edges out of $r-1$ in
$G_{r-1}$ are stored in a global array $h$ of size $N$. 

\bigskip

By Theorem \ref{t_window},
\[
r=\min\left\{  n\in(s+m:N-M+m]:d_{s}^{\min}\left(  \delta_{s}\left(
m,n\right)  ,N\right)  \leq M\right\}
\]
The downward monotonicity of $\delta_{s}\left(  m,n\right)  $ and $d_{s}%
^{\min}\left(  \delta_{s}\left(  m,n\right)  ,N\right)  $ in $n\in(s+m:N-M+m]$
enables the discovery of $r$ by a combination of \emph{exponential search} and
\emph{binary search}. Throughout the search, an integer search interval
$(n^{\prime}:n^{\prime\prime}]$ containing $r$ is maintained; initially,
$n^{\prime}=s+m$ and $n^{\prime\prime}=N-M+m$. A basic sampling operation is
to generate a sample $\lambda=\delta_{s}\left(  m,n\right)  $ for some
candidate $n$ in the search interval, compute $M^{\prime}=d_{s}^{\min}\left(
\lambda,N\right)  $, and compare $M^{\prime}$ against $M$ as follows:

\begin{itemize}
\item If $M^{\prime}=M$, then $\lambda$ is returned.

\item If $M^{\prime}>M$, then $n<r$ hence $n^{\prime}$ is lifted to $n$.

\item If $M^{\prime}<M$, then $n\geq r$ hence $n^{\prime\prime}$ is reduced to
$n$.
\end{itemize}

\noindent The procedure \textbf{Probe}$\left(  s,m\right)  $ runs in three
phases, exponential search, binary search, and contraction, which are
elaborated below.

\bigskip

The exponential search is outlined in Table \ref{tab_ES}. It takes
$k:=\left\lceil \log\left(  r-s-m+1\right)  \right\rceil $ iterations to
either hit a member $\lambda\in\Lambda_{s}\left(  M,N\right)  $ or reach
\begin{align*}
n^{\prime} &  =\left(  s+m-1\right)  +2^{k-1},\\
n^{\prime\prime} &  =\min\left\{  \left(  s+m-1\right)  +2^{k},N-M+m\right\}
.
\end{align*}
For each $j\in\left[  k\right]  $, the $j$-th iteration selects the candidate
\[
n=\min\left\{  \left(  s+m-1\right)  +2^{j},N-M+m\right\}  ,
\]
invokes \textbf{PBF}$\left(  s,m,n\right)  $ on $G_{s}$ to compute $f\left(
i\right)  =f_{s}\left(  m,i\right)  $ for $s+m\leq i\leq n$ and $\overline
{f}\left(  i\right)  =f_{s}\left(  m+1,i\right)  $ for $s+m+1\leq i\leq n$,
and completes the sampling operation on $\lambda=\delta_{s}\left(  m,n\right)
=\overline{f}\left(  n\right)  -f\left(  n\right)  $. If the sampling
operation reduces $n^{\prime\prime}$ to $n$, then the binary search follows,
inheriting $f\left(  i\right)  =f_{s}\left(  m,i\right)  $ for $s+m\leq i\leq
n^{\prime\prime}$ and $\overline{f}\left(  i\right)  =f_{s}\left(
m+1,i\right)  $ for $s+m+1\leq i\leq n^{\prime\prime}$. %

\begin{table}[htbp] \centering
\begin{tabular}
[c]{|l|}\hline
// exponential search:\\\hline\hline
$n^{\prime}\leftarrow s+m$, $n^{\prime\prime}\leftarrow N-M+m$, $l\leftarrow
1$;\\
repeat\\
\quad$n\leftarrow\min\left\{  n^{\prime}+l,n^{\prime\prime}\right\}  $;\\
\quad\textbf{PBF}$\left(  s,m,n\right)  $;\\
\quad$\lambda\leftarrow\overline{f}\left(  n\right)  -f\left(  n\right)  $;\\
\quad$M^{\prime}\leftarrow d_{s}^{\min}\left(  \lambda,N\right)  $;\\
\quad if $M^{\prime}=M$ then return $\lambda$;\\
\quad if $M^{\prime}>M$ then $n^{\prime}\leftarrow n$, $l\leftarrow2l$;\\
until $M^{\prime}<M$;\\
$n^{\prime\prime}\leftarrow n$;\\\hline
\end{tabular}
\caption{Outline of exponential search.}\label{tab_ES}%
\end{table}

\bigskip

The binary search is outlined in Table \ref{tab_BS}. As long as $n^{\prime
\prime}-n^{\prime}>1$, a binary-search iteration chooses the median
$n:=\left\lceil \left(  n^{\prime}+n^{\prime\prime}\right)  /2\right\rceil $ and completes the sampling operation on $\lambda=\delta_{s}\left(  m,n\right)
=\overline{f}\left(  n\right)  -f\left(  n\right)  $. \noindent When
$n^{\prime\prime}-n^{\prime}=1$, $r=n^{\prime\prime}$ and the membership of
$\lambda=\delta_{s}\left(  m,r\right)  =\overline{f}\left(  r\right)
-f\left(  r\right)  $ in $\Lambda_{s}\left(  M,N\right)  $ is tested by
comparing $M^{\prime\prime}=d_{s}^{\max}\left(  \lambda,N\right)  $ against
$M$. If $M^{\prime\prime}\geq M$ then $\delta_{s}\left(  m,r\right)
\in\Lambda_{s}\left(  M,N\right)  $ and is thus returned. Otherwise,
\[
\delta_{s}\left(  m,r\right)  <\delta_{s}\left(  M-1,N\right)  \leq\delta
_{s}\left(  M,N\right)  <\delta_{s}\left(  m,r-1\right)  .
\]
and the contraction phase follows. %

\begin{table}[htbp] \centering
\begin{tabular}
[c]{|l|}\hline
// binary search:\\\hline\hline
while $n^{\prime\prime}-n^{\prime}>1$ do\\
\quad$n\leftarrow\left\lceil \left(  n^{\prime}+n^{\prime\prime}\right)
/2\right\rceil $, $\lambda\leftarrow\overline{f}\left(  n\right)  -f\left(
n\right)  $;\\
\quad$M^{\prime}\leftarrow d_{s}^{\min}\left(  \lambda,N\right)  $;\\
\quad if $M^{\prime}=M$ then return $\lambda$;\\
\quad if $M^{\prime}>M$ then $n^{\prime}\leftarrow n$ else $n^{\prime\prime
}\leftarrow n$;\\
$r\leftarrow n^{\prime\prime},\lambda\leftarrow\overline{f}\left(  r\right)
-f\left(  r\right)  $;\\
$M^{\prime\prime}\leftarrow d_{s}^{\max}\left(  \lambda,N\right)  $;\\
if $M^{\prime\prime}=M$ then return $\lambda$;\\\hline
\end{tabular}
\caption{Outline of binary search.}\label{tab_BS}%
\end{table}%

\bigskip

The contraction phase is outlined in Table \ref{tab_Contract}. It first
computes
\[
h\left(  j\right)  =\min_{s+m\leq i<r}\left[  f\left(  i\right)  +c_{s}\left(
i,j\right)  \right]
\]
for each $j\in\left[  r:N\right]  $. Since $f\left(  i\right)  +c_{s}\left(
i,j\right)  $ is submodular in $\left(  i,j\right)  $ on the lattice $\left[
s+m:r-1\right]  \times\left[  r:N\right]  $,  $h\left(  j\right)  $ for
$j\in\left[  r:N\right]  $ can be computed by the SWAWK\ algorithm.
Subsequently, $s$ is overwritten with $r-1$ and $M$ is decreased by $m$.%

\begin{table}[htbp] \centering
\begin{tabular}
[c]{|l|}\hline
// contraction:\\\hline\hline
for $j=r$ to $N$ do //SMAWK\\
\quad$h\left(  j\right)  \leftarrow\min_{s+m\leq i<r}\left[  f\left(
i\right)  +c_{s}\left(  i,j\right)  \right]  $;\\
$s\leftarrow r-1$, $M\leftarrow M-m$;\\\hline
\end{tabular}
\caption{Outline of the contraction phase.}\label{tab_Contract}%
\end{table}%

\bigskip

In the remaining of this section, we derive the time and space complexity of
the procedure \textbf{Probe}$\left(  s,m\right)  $.

\bigskip

\begin{lemma}
\label{l_HoC} The procedure \textbf{Probe}$\left(  s,m\right)  $ runs in%
\[
O\left(  m\left(  r-s-m+1\right)  +\left(  N-s+1\right)  \log\left(
r-s-m+1\right)  \right)  .
\]
time and $O\left(  N\right)  $ space.
\end{lemma}

\bigskip

\begin{proof}
The worst-case running time occurs when the contraction phase is reached. The
contraction phase has $O\left(  N-s+1\right)  $ running time. The total
running time of the other two phases is dominated by the invocations of
\textbf{PBF}$\left(  s,m,n\right)  $ and the computations of $d_{s}^{\min
}\left(  \lambda,N\right)  $ or $d_{s}^{\max}\left(  \lambda,N\right)  $. We
show that the former's total running time is
\[
O\left(  m\left(  r-s-m+1\right)  \right)
\]
and the latter's total running time is
\[
O\left(  \left(  N-s+1\right)  \log\left(  r-s-m+1\right)  \right)  ,
\]
from which the lemma follows.

Let
\[
k=\left\lceil \log\left(  r-s-m+1\right)  \right\rceil .
\]
Then the exponential search takes $k$ iterations. In the $j$-th iteration with
$j\in\left[  k\right]  $,
\[
n=\min\left\{  \left(  s+m-1\right)  +2^{j},N-M+m\right\}  ;
\]
hence the invocation \textbf{PBF}$\left(  s,m,n\right)  $ has running time
\[
O\left(  m\left(  n-s+1-m\right)  \right)  =O\left(  m2^{j}\right)  .
\]
Since
\[
m\sum_{j=1}^{k}2^{j}=2m\left(  2^{k}-1\right)  <4m\left(  r-s-m+1\right)  ,
\]
the invocations of \textbf{PBF}$\left(  s,m,n\right)  $ in the exponential
search is
\[
O\left(  m\left(  r-s-m+1\right)  \right)  .
\]

The exponential search takes $k$ computations of $d_{s}^{\min}\left(
\lambda,N\right)  $. At the beginning of the binary search,
\begin{align*}
n^{\prime} &  =\left(  s+m-1\right)  +2^{k-1},\\
n^{\prime\prime} &  \leq\left(  s+m-1\right)  +2^{k}.
\end{align*}
As $n^{\prime\prime}-n^{\prime}\leq2^{k-1}$, the binary search takes at most
$k-1$ iterations, hence has at most $k-1$ computations of $d_{s}^{\min}\left(
\lambda,N\right)  $ and one computation of $d_{s}^{\max}\left(  \lambda
,N\right)  $. Thus, there at most $2k$ computations of $d_{s}^{\min}\left(
\lambda,N\right)  $ or $d_{s}^{\max}\left(  \lambda,N\right)  $ in total, and
each computation takes $O\left(  N-s+1\right)  $ time. So, the total time by
the computations of $d_{s}^{\min}\left(  \lambda,N\right)  $ or $d_{s}^{\max
}\left(  \lambda,N\right)  $ is
\[
O\left(  \left(  N-s+1\right)  \log\left(  r-s-m+1\right)  \right)  .
\]

The linear space complexity of \textbf{Probe}$\left(  s,m\right)  $ follows
from that both \textbf{PBF}$\left(  s,m,n\right)  $ and the computation of
$d_{s}^{\min}\left(  \lambda,N\right)  $ or $d_{s}^{\max}\left(
\lambda,N\right)  $ require linear space.
\end{proof}

\bigskip

\section{The Contract-and-Conquer Algorithm}

\label{s_CC}

\bigskip

Suppose $G_{1}=G$ is a Monge DAG on $\left[  N\right]  $ with edge-length
function $c_{1}=c$, and $2\leq M\leq N-2$. Denote $\Lambda^{\ast}:=\Lambda
_{1}\left(  M,N\right)  $. Computing a path in $\mathcal{P}_{1}^{\ast}\left(
M,N\right)  $ is reducible in \emph{linear} time and space to finding a member
$\lambda\in\Lambda^{\ast}$. In this section, we present a contract-and-conquer
(C\&C) algorithm for finding a member $\lambda\in\Lambda^{\ast}$ in linear
space and $O\left(  \sqrt{NM\left(  N-M\right)  \log\left(  N-M\right)
}\right)  $ time. 

\bigskip

If \ $M\left(  N-M\right)  \leq4N\log\left(  N-M\right)  $, then the C\&C
algorithm simply invokes \textbf{PBF}$\left(  1,M,N\right)  $ on $G_{1}$ and
returns $\lambda=\delta_{1}\left(  M,N\right)  $ in linear space and
\[
O\left(  M\left(  N-M\right)  \right)  =O\left(  \sqrt{NM\left(  N-M\right)
\log\left(  N-M\right)  }\right)
\]
time. In the remaining of this section, we assume
\[
M\left(  N-M\right)  >4N\log\left(  N-M\right)  .
\]
Then $16<M<N-16$. Indeed, $N-M>16$ follows from
\[
\left(  N-M\right)  /\log\left(  N-M\right)  >4N/M>4,
\]
which further implies
\[
M>4\frac{N}{N-M}\log\left(  N-M\right)  >4\log\left(  N-M\right)  >16.
\]

\bigskip

Let
\begin{equation}
K:=\left\lceil \sqrt{\frac{M\left(  N-M\right)  }{N\log\left(  N-M\right)  }%
}\right\rceil .\label{eq_stages}%
\end{equation}
Then $2<K<\sqrt{M}$ as
\[
2<K\leq\left\lceil \sqrt{\frac{M}{\log\left(  N-M\right)  }}\right\rceil
\leq\left\lceil \sqrt{M}/2\right\rceil <\sqrt{M}.
\]
Hence $K/M>\sqrt{M}>4$. Denote $K^{\prime}:=K-M\operatorname{mod}K$, and
partition $M$ \emph{evenly} into $K$ integers $m_{1},m_{2},\cdots,m_{K}$
where
\[
m_{k}=\left\{
\begin{array}
[c]{cc}%
\left\lfloor M/K\right\rfloor , & \text{if }1\leq k\leq K^{\prime};\\
\left\lceil M/K\right\rceil , & \text{if }K^{\prime}<k\leq K.
\end{array}
\right.
\]
Then, $m_{k}\geq4$ for each $1\leq k\leq K$. 

\bigskip

The C\&C algorithm \emph{implicitly} maintains a Monge DAG\ $G_{s}$ on
$\left[  s:N\right]  $ with edge length function $c_{s}$ for some $s\geq1$.
Initially, $s=1$. Each $G_{s}$ with $s>1$ is generated from $G_{1}$ by
successive contractions and is \emph{explicitly} represented by a global
array $h$ of size $N$ such that $h\left(  j\right)  =c_{s}\left(  s,j\right)
$ for each $j\in\left[  s+1:N\right]  $. The C\&C algorithm proceeds in at
most $K$ successive stages. At the beginning of the stage $k\leq K$, the
following invariant properties are maintained:  

\begin{itemize}
\item $M=\sum_{i=k}^{K}m_{i}\leq N-s-1$; 

\item $\Lambda_{s}\left(  M,N\right)  $ in $G_{s}$ is $\Lambda^{\ast}$. 
\end{itemize}

\noindent These properties hold trivially for the first stage $k=1$ where
$s=1$. The stage $k$ runs as follows depending on whether $k=K$ or not.

\bigskip

Suppose $k<K$. As
\begin{align*}
4  & \leq m_{k}=M-\sum_{i=k+1}^{K}m_{i}\leq M-4,\\
4  & \leq M\leq N-s-1,
\end{align*}
the stage $k$ invokes \textbf{Probe}$\left(  s,m_{k}\right)  $ on $G_{s}$. If
the invocation of \textbf{Probe}$\left(  s,m_{k}\right)  $ finds a member
$\lambda\in\Lambda_{s}\left(  M,N\right)  =\Lambda^{\ast}$, then the C\&C
algorithm \emph{terminates} with the output $\lambda$. Otherwise, the
invocation of \textbf{Probe}$\left(  s,m_{k}\right)  $ \emph{contracts}
$G_{s}$ by updating $s$ and $h$ and reduces $M$ by $m_{k}$. After the
contraction, 

\begin{itemize}
\item $M=\sum_{i=k+1}^{K}m_{i}\leq N-s-1$; 

\item $\Lambda_{s}\left(  M,N\right)  $ in $G_{s}$ is $\Lambda^{\ast}$ by
Theorem \ref{t_probe}. 
\end{itemize}

\noindent The C\&C algorithm then moves on to the stage $k+1$. 

\bigskip

Suppose $k=K$. Then the (last) stage $K$ simply invokes \textbf{PBF}$\left(
s,M,N\right)  $ on $G_{s}$, computes $\lambda=\delta_{s}\left(  M,N\right)  $,
and \emph{terminates} with the output $\lambda$.  

\bigskip

The data structures needed by the C\&C algorithm are just a few global arrays
indexed of size $N$. The array $h$ is used for representing the graph $G_{s}$.
Two arrays $f$ and $\overline{f}$ are used by the procedure \textbf{PBF}%
$\left(  s,m,n\right)  $. The computation of $d_{s}^{\min}\left(
\lambda,N\right)  $ or $d_{s}^{\max}\left(  \lambda,N\right)  $ and the
invocation of SMAWK algorithm require linear working space. We remark that
there is no need to explicitly maintain the sequence $m_{1},m_{2},\cdots
,m_{K}$. They can be easily derived from the stage number $k$ and $K^{\prime
}=K-M\operatorname{mod}K$. Thus, the C\&C algorithm has linear space
complexity.  

\bigskip

Next, we derive the time complexity of the C\&C algorithm.

\bigskip

\begin{theorem}
The C\&C algorithm has time complexity $O\left(  \sqrt{NM\left(  N-M\right)
\log\left(  N-M\right)  }\right)  $. 
\end{theorem}

\bigskip

\begin{proof}
Clearly, the worst-case running time occurs when the stage $K$ is reached. For
each $1\leq k\leq k$, let $s_{k}$ be the root $s$ at the beginning of the
stage $k$. Then, $s_{1}=1$; and for each $k<K$, $s_{k+1}+1$ is exactly the
pivot $r$ in stage $k$. Thus, for each $k<K$, $s_{k+1}+1\geq s_{k}+m_{k}+1$
implying that $s_{k+1}\geq s_{k}+m_{k}$. At the beginning of the last stage
$K$, $m_{K}\leq N-s_{K}-1$; hence
\begin{equation}
s_{K}+m_{K}+1\leq N.\label{eq_laststage}%
\end{equation}

By Lemma \ref{l_HoC}, each stage $k<K$ has running time
\begin{align*}
&  O\left(  m_{k}\left(  s_{k+1}-s_{k}-m_{k}+2\right)  \right)  +\\
&  O\left(  \left(  N-s_{k}+1\right)  \log\left(  s_{k+1}-s_{k}-m_{k}%
+2\right)  \right)  .
\end{align*}
The stage $K$ has running time
\[
O\left(  m_{K}\left(  N-s_{K}-m_{K}+1\right)  \right)  .
\]
So, the total running time is
\begin{align*}
&  O\left(  \sum_{k=1}^{K-1}m_{k}\left(  s_{k+1}-s_{k}-m_{k}+2\right)
+m_{K}\left(  N-s_{K}-m_{K}+1\right)  \right)  \\
&  +O\left(  \sum_{k=1}^{K-1}\left(  N-s_{k}+1\right)  \log\left(
s_{k+1}-s_{k}-m_{k}+2\right)  \right)
\end{align*}
We show the two sums in the big-$O$ notations are bounded respectively by
\begin{align*}
& \frac{M}{K}\left(  N-M\right)  +N+M,\\
& N\left(  K-1\right)  \log\left(  n-M\right)  .
\end{align*}

By the definition of $m_{k}$ for $k\in\left[  K\right]  $, we have
\begin{align*}
&  \sum_{k=1}^{K-1}m_{k}\left(  s_{k+1}-s_{k}-m_{k}+2\right)  +m_{K}\left(
N-s_{K}-m_{K}+1\right)  \\
&  <\sum_{k=1}^{K-1}m_{k}\left(  s_{k+1}-s_{k}-m_{k}\right)  +m_{K}\left(
N-s_{K}-m_{K}\right)  +2M\\
&  \leq\left\lceil \frac{M}{K}\right\rceil \sum_{k=1}^{K-1}\left(
s_{k+1}-s_{k}-m_{k}\right)  +\left\lceil \frac{M}{K}\right\rceil \left(
N-s_{K}-m_{K}\right)  +2M\\
&  =\left\lceil \frac{M}{K}\right\rceil \left(  N-M-1\right)  +2M\\
&  <\frac{M}{K}\left(  N-M\right)  +N+M.
\end{align*}

By the concavity of the logarithm function and the inequality in equation
(\ref{eq_laststage}), we have
\begin{align*}
&  \sum_{k=1}^{K-1}\left(  N-s_{k}+1\right)  \log\left(  s_{k+1}-s_{k}%
-m_{k}+2\right)  \\
&  <N\sum_{k=1}^{K-1}\log\left(  s_{k+1}-s_{k}-m_{k}+2\right)  \\
&  \leq N\left(  K-1\right)  \log\left(  \frac{\sum_{i=1}^{K-1}\left(
s_{k+1}-s_{k}-m_{k}\right)  }{K-1}+2\right)  \\
&  =N\left(  K-1\right)  \log\left(  \frac{s_{K}+m_{K}-M-1}{K-1}+2\right)  \\
&  \leq N\left(  K-1\right)  \log\left(  \frac{N-M-2}{K-1}+2\right)  \\
&  \leq N\left(  K-1\right)  \log\left(  n-M\right)  .
\end{align*}

Finally, by the choice of $K$ in equation (\ref{eq_stages}),
\[
\frac{M}{K}\left(  N-M\right)  \leq\frac{M\left(  N-M\right)}{\sqrt{\frac{M\left(  N-M\right)
}{N\log\left(  N-M\right)  }}}=\sqrt{NM\left(  N-M\right)  \log\left(
N-M\right)  }.
\]
and
\begin{align*}
&  N\left(  K-1\right)  \log\left(  n-M\right)  \leq N\sqrt{\frac{M\left(
N-M\right)  }{N\log\left(  N-M\right)  }}\log\left(  N-M\right)  \\
&  =\sqrt{NM\left(  N-M\right)  \log\left(  N-M\right)  }.
\end{align*}
Therefore, the theorem holds.
\end{proof}

\bigskip

The rationale for the choice of $K$, the number of stages, given by equation
(\ref{eq_stages}) is now clear from the above proof. This choice is to strike
a balance between the invocations of \textbf{PBF}$\left(  s,m,n\right)  $ and
the computations of $d_{s}^{\min}\left(  \lambda,N\right)  $ or $d_{s}^{\max
}\left(  \lambda,N\right)  $. 

\bigskip

\section{Conclusion}

\label{s_conclude}

\bigskip

For the $O\left(  poly\left(  \log N\right)  \right)  $ and $N-O\left(
poly\left(  \log N\right)  \right)  $ regimes of $M$, the C\&C algorithm has
running time $O\left(  N\cdot poly\left(  \log N\right)  \right)  $. It
remains open whether there is $O\left(  N\cdot poly\left(  \log N\right)
\right)  $-time algorithm in the regime of $M$ where both $M$ and $N-M$ are at
least $\Omega\left(  poly\left(  \log N\right)  \right)  $. It seems quite
possible that with some additional cleverness our algorithm could be made to
run faster. Note that the membership test of $\lambda$ in $\Lambda^{\ast
}=\Lambda_{1}\left(  M,N\right)  $ via $d_{s}^{\min}\left(  \lambda,N\right)
$ and $d_{s}^{\max}\left(  \lambda,N\right)  $ in $G_{s}\left(  \lambda
\right)  $ can also be done via $d_{1}^{\min}\left(  \lambda,N\right)  $ and
$d_{1}^{\max}\left(  \lambda,N\right)  $ in $G_{1}\left(  \lambda\right)  $,
and the same asymptotic upper bound $O\left(  \sqrt{NM\left(  N-M\right)
\log\left(  N-M\right)  }\right)  $ on the total running time is still valid.
Each time we compute $d_{1}^{\min}\left(  \lambda,N\right)  $ or $d_{1}^{\max
}\left(  \lambda,N\right)  $, we actually produce the parametric shortest-path
tree $T_{1}^{\min}\left(  \lambda,N\right)  $ or $T_{1}^{\max}\left(
\lambda,N\right)  $. Our current algorithm utilizes only the value
$d_{1}^{\min}\left(  \lambda,N\right)  $ or $d_{1}^{\max}\left(
\lambda,N\right)  $ and throws away all the information about the
shortest-path tree. Possibly some information of the shortest-path trees could
be reused to achieve an improved running time. All strongly polynomial
algorithms seek a member of $\Lambda^{\ast}$ from the finite candidate pool of
$\delta_{s}\left(  m,n\right)  $ for $\left(  m,n\right)  \in\mathcal{L}_{s}$.
One may expand the finite candidate pool with extra candidates which can be
generated easily and lead to faster progress. Ultimately, the disparate
dependence on $M$ of the running times of different algorithms suggests that a
faster \textquotedblleft hybrid\textquotedblright\ algorithm may be composed
in an adaptive manner.

\bigskip


\begin{thebibliography}{99}                                                                                               %


\bibitem {AKM+87}A. Aggarwal, M. Klawe, S. Moran, P. Shor, and R. Wilber,
Geometric applications of a matrix-searching algorithm, \emph{Algorithmica} 2:
195--208, 1987.

\bibitem {AST94}A. Aggarwal, B. Schieber, and T. Tokuyama, Finding a minimum
weight $K$-link path in graphs with Monge property and applications,
\emph{Discrete Comput. Geometry} 12: 263-280, 1994.

\bibitem {BLP92}W. Bein, L. Larmore, and J. Park, The $d$-edge shortest-path
problem for a Monge graph, Preprint, 1992.

\bibitem {GP90}Z. Galil and K. Park, A linear-time algorithm for concave
one-dimensional dynamic programming, \emph{Inform. Process. Lett.} 33 (6):
309-311. 1990.

\bibitem {Kla89}M. Klawe, A simple linear time algorithm for concave
one-dimensional dynamic programming,\ \emph{Technical Report 89-16},
University of British Columbia, Vancouver, 1989.

\bibitem {LS91}L. Larmore and B. Schieber, On-line dynamic programming with
applications to the prediction of RNA secondary structure, \emph{Journal of
Algorithms} 12 (3): 490-515, 1991.

\bibitem {Meg83}N. Megiddo, Applying parallel computation algorithms in the
design of serial algorithms, \emph{J. Assoc. Comput. Mach.} 30: 852-865, 1983.

\bibitem {Sch98}B. Schieber, Computing a minimum weight $k$-link path in
graphs with the concave monge property, \emph{Journal of Algorithms} 29:
204-222, 1998.

\bibitem {Top11}D. Topkis. \emph{Supermodularity and complementarity},
Princeton University Press, 2011.

\bibitem {Wil88}R. Wilber, The concave least weight subsequence problem
revisited, \emph{Journal of Algorithms} 9: 418--425, 1988.
\end{thebibliography}
\end{document}